\documentclass[aps,epsf,showpacs]{revtex4}

\begin{document}
\title{Dirac fermions on a disclinated flexible surface}

\author{E.A. Kochetov}\email{kochetov@theor.jinr.ru}

\author{V.A. Osipov}\email{osipov@theor.jinr.ru}

\affiliation{Bogoliubov Laboratory of Theoretical Physics, Joint
Institute for Nuclear Research, 141980 Dubna, Moscow region,
Russia}

\begin{abstract}
A self-consisting gauge-theory approach to describe Dirac fermions
on flexible surfaces with a disclination is formulated. The elastic
surfaces are considered as embeddings into $R^3$ and a disclination
is incorporated through a topologically nontrivial gauge field of
the local $SO(3)$ group which generates the metric with conical
singularity. A smoothing of the conical singularity on flexible
surfaces is naturally accounted for by regarding the upper half of
two-sheet hyperboloid as an elasticity-induced embedding. The
availability of the zero-mode solution to the Dirac equation is
analyzed.
\end{abstract}


\maketitle

\section{Introduction}

Is is now generally accepted that the low-lying electronic states in
graphene can be accurately described by two-dimensional massless
Dirac fermions~\cite{review}. In experiment, multiform graphene
structures were observed thus stimulating studies of Dirac fermions
on curved graphene sheets (see, e.g.,~\cite{cortijo,pachos}). This
problem is markedly complicated when the curvature itself is
generated by topological defects like disclinations. Indeed, a
disclination is known in elasticity theory as a line defect which
can be produced by "cut and glue" Volterra process, namely, by
inserting or removing a wedge of material with the following gluing
of the dihedral sides. This immediately generates additional large
elastic strains inside the crystal. For flexible membranes, however,
there is a chance to screen out the strain field by buckling into a
cone. The problem thus reduces to coupling Dirac spinors to a
topologically non-trivial curved background.

According to Volterra process disclination can be considered as a
conical singularity like strings in cosmology. The relevant background
is the curved spacetime where all the
curvature is concentrated at the apex of the cone. The metric of this
$2D$ space in polar coordinates is written as
\begin{equation}
ds^2 =dr^2+\alpha^2r^2d\varphi^2. \label{metric}
\end{equation}
Here the parameter $\alpha$ is related to the angular sector that
is removed or inserted to form the defect. In this case, any
attempt to build a closed loop around the disclination line will
result in a closure failure. The deficit angle is equal to
2$\pi\alpha$ with $\alpha=1-\nu$ where $\nu$ is the Frank index,
the basic topological characteristic of the disclination. The
positive sign of $\nu$ corresponds to the removing of a sector. In
this case the space has positive curvature. Correspondingly, for
negative $\nu$ one has a cone of negative curvature. Eventually, the
problem reduces to a Dirac equation in the curved
spacetime.

In spite of the elegant form of this approach, there is yet an
important open question concerning the so-called core region of the
defect. To the best of our knowledge, for the first time this
problem was raised in cosmological models~\cite{allen1,allen2} where
long-range effects of cosmic string cores were studied. In geometric
theory of defects, an influence of a disclination core on the
localization of electrons and holes was investigated
in~\cite{ribeiro}. In both cases, the tip of the conical singularity
is replaced by a smooth cap while at large distances a typical cone
with the deficit angle 2$\pi\alpha$ emerges. In cosmological models
the curvature of an infinite strait string is confined within a
cylinder of a small radius $a$ (the core radius) that possesses a
direct physical meaning: it characterizes the interior structure of
the string. Accordingly, the relevant metric can be taken in the
form
\begin{equation}
ds^2 = dt^2+dz^2+ P^2(r/a)dr^2+r^2d\varphi^2, \label{coremetric}
\end{equation}
where the range of the angular coordinate is $\varphi\in [0, 2\pi\alpha$) and
$P(r/a)$ is a smooth monotonic function satisfying the conditions
\begin{equation}
\lim_{r/a\rightarrow 0} P(r/a) = \nu, \qquad  P(r/a)=1, \quad r > a.
\label{P}
\end{equation}
For example, in~\cite{ribeiro}
the so-called flower-pot model was considered when the curvature of the disclinated
media is concentrated on a ring of radius $a$, which results in the formation of a "seam"
on the cylinder.

It should be stressed that
this approach is of interest in the description of linear defects
with a certain interior structure (finite thickness of a string).
However, the situation changes drastically for a disclination on an elastic $2D$ surface.
First of all, in this case there is no parameter (similar to $a$) that fixes a relevant
short-range length scale: a disclination is a point defect.
Second, the specificity
of 2D elastic surfaces lies in that they may change both their
intrinsic and extrinsic geometries. For example, the creation of a disclination
in a nonstretchable membrane by using the "cut and glue" process will result
in a true cone. In reality, however, the membranes are flexible and the cone
apex will be smoothed due to finite elasticity. Elastic deformations are by definition smooth
deformations.
A cone cannot be smoothly evolved into a desired surface with a smoothed apex
simply because of the fact that
a cone is not a manifold. If one ignores this and just try to formally carry over
the $P$-type smearing procedure used for strings to
elastisity theory,
one will inevitably run into
a problem of fixing boundary conditions at $r=a$. In the present case those conditions are purely
artificial and possess
no direct physical meaning (see also discussion in~\cite{nelson}).
In other words, it is not a straightforward matter
to incorporate self-consistently the information about the core region
within the geometrical approach in $2D$ elastic theory with defects.

In this paper, we attempt to develop a variant of the
self-consistent gauge-theory approach
to take into account both the smoothed apex and the
topological characteristic of the defect. Actually, a part of our
program was already realized in~\cite{jpa99}. The model developed
there allows us to describe disclinations on arbitrary
elastic surfaces. It includes Riemannian surfaces that may change
their geometry under deformations. The local gauge field to describe
disclinations on an elastic surface emerges as a gauge field of the
local $SO(3)$ group of the local rotations of $R^3$. This ensures the
local rotational invariance of the elasticity action. In spite of the
fact that we consider $2D$ manifolds, the group $SO(3)$ appears
rather than the $SO(2)$ one because of the fact that we formulate
our theory in terms of the embeddings of a $2D$ surface into $R^3$.
By construction the local gauge field affects the underlying metric.
Within the linear scheme the model recovers the von Karman equations
for membranes with a disclination-induced source being generated by
gauge fields. For a single disclination on an arbitrary surface a
covariant generalization of these equations is obtained.

The dynamical variables of our theory are the embeddings
$R^i(x^1,x^2)$ and gauge fields $W_{\mu}^i$ to be determined
self-consistently (indices $\mu,\nu,..=1,2$ are tangent to the
surface, whereas $i,j,..=1,2,3$ run over the basis of $R^3$). As
the outcome, the induced metric
$$g_{\mu\nu}(W)=\nabla_\mu\vec R \nabla_\nu\vec R, \qquad \nabla_\mu=\partial_\mu+[\vec W_{\mu},...]$$
emerges. Explicitly,
\begin{eqnarray}
g_{\mu\nu}(W)&=&
\partial_{\mu}\vec R\cdot
\partial_{\nu}\vec R+
\partial_{\mu}\vec R[\vec W_{\nu},\vec R] +
\partial_{\nu}\vec R[\vec W_{\mu},\vec R] \nonumber \\
&+&(\vec W_{\mu} \vec W_{\nu})\vec R^2 - (\vec W_{\mu}\vec R)(\vec
W_{\nu}\vec R). \label{2.0}\end{eqnarray} In general, the
dynamical fields $R^i$ and $W_a^i$ couple to each other. However,
in the linear in elastic field approximation the gauge field can
be considered as an external field~\cite{jpa99}.

\section{Dirac fermions on a manifold with a dynamically induced metric}

The important issue is how a non-trivial gauge potential can
explicitly be incorporated into the theory to self-consistently
describe disclination defects on an elastic surface with fermions.
To incorporate Dirac fermions we observe that the topologically
nontrivial gauge field reasserts itself in the Dirac equation as a
topologically nontrivial $SO(2)$ piece of the spin
connection~\cite{jetp}. That part of the connection carries a
topologically nontrivial flux that does not depend on smooth
continuous changes of the underlying metric due to small elastic
deformations.

To incorporate fermions on the $2D$ curved background $(\Sigma,\, g_{\mu\nu}(W))$
we need a set of orthonormal frames $\{e_{\alpha}(W)\}$ which yield the same metric,
$g_{\mu\nu}(W)$, related to each other by the local $SO(2)$ rotation,
$$e_{\alpha}\to e'_{\alpha}={\Lambda}_{\alpha}^{\beta}e_{\beta},\quad
{\Lambda}_{\alpha}^{\beta}\in SO(2).$$
It then follows that
$g_{\mu\nu} = e^{\alpha}_{\mu}e^{\beta}_{\nu} \delta_{\alpha
\beta}$ where $e_{\alpha}^{\mu}$ is the zweibein, with the
orthonormal frame indices being $\alpha,\beta=\{1,2\}$, and
coordinate indices $\mu,\nu=\{1,2\}$ (from now on we drop an explicit
$W$-dependance of the metric). As usual, to ensure that
physical observables be independent of a particular choice of the
zweinbein fields, a local $so(2)$--valued gauge field
$\omega_{\mu}$ is to be introduced. The gauge field of the local
$SO(2)$ group is referred to as a spin connection. For the theory to be
self-consistent, zweinbein fields must be chosen to be covariantly
constant \cite{witten}:
$$\partial_{\mu}e^{\alpha}_{\nu}
-\Gamma^{\lambda}_{\mu\nu}e^{\alpha}_{\lambda}+(\omega_{\mu})^{\alpha}_{\beta}
e^{\beta}_{\nu}=0,$$ which determines the spin connection
coefficients explicitly
\begin{equation}
(\omega_{\mu})^{\alpha\beta}= e_{\nu}^{\alpha}D_{\mu}e^{\beta\nu},\quad D_{\mu}=\partial_{\mu}+\Gamma_{\mu},
\label{2.1}\end{equation}
with $\Gamma_{\mu}$ being the Levi-Civita connection.
The Dirac equation on a surface $(\Sigma,\, g_{\mu\nu}(W))$ is written as
\begin{equation}
i\gamma^{\alpha}e_{\alpha}^{\ \mu}(\partial_{\mu}+\Omega_{\mu})\psi=E\psi,
 \label{2.2}\end{equation}
with
\begin{equation}
\Omega_{\mu}=\frac{1}{8}\omega^{\alpha\ \beta}_{\ \mu}
[\gamma_{\alpha},\gamma_{\beta}] \label{2.3}\end{equation} being
the spin connection in the spinor representation.

Let us consider first a plane which can be bent but can not be stretched.
A single disclination can be inserted in this plane by using the "cut and glue"
Volterra process. Obviously, the resulting surface is nothing else but a cone.
In our description, we start from a $2D$ flat metric disturbed by a disclination
defect.

In the polar coordinates
$(r,\varphi)\in R^2$ a plane can be regarded as an embedding
$$(r,\varphi)\to
(r\cos\varphi, \ r\sin\varphi, 0), \quad 0<r<1,\,
0\le\varphi<2\pi.$$ The disclination defect is placed at the origin
and is described by the gauge field $W_{\mu}^{i=1,2}=0$ and
$W_{\mu}^{i=3}=W_{\mu}$, where in the polar coordinates~\cite{jpa99}
\begin{equation}
W_r=0,\,\, W_{\varphi}=\nu.
\label{0.0}\end{equation}

Notice that for any counter $C$ encircling the origin one has
\begin{equation}
\oint_{C} W=2\pi\nu\ne 0. \label{0.1}\end{equation} Since the
counter integral in (\ref{0.1}) is a gauge invariant quantity, the
field $W_{\mu}$ cannot be gauged away to zero due to the topological
obstruction. This is why that field is referred to as a
topologically non-trivial one. A physically observable quantity
associated with that gauge field is a nonzero flux, $\Phi=2\pi\nu$,
through an area bounded by the counter $C$. It does not depend on
small continuous deformations of that area. This flux instead
characterizes the gauge potential globally: it determines the first
Chern characteristic class the gauge potential $W$ belongs to. An
electron encircling the origin naturally acquires a topological
phase associated with that nontrivial flux: the Aharonov-Bohm phase
which distinguishes the gauge potential $W$ from a trivial one.

The components of the induced metric (\ref{2.0}) can be easily read
off
\begin{equation} g_{rr}= 1, \qquad
g_{\varphi\varphi}=\alpha^2r^2, \qquad g_{r\varphi}=g_{\varphi r}=0,
\label{eq:3.1}\end{equation} where $\alpha=1-\nu.$ Evidently, this
is a metric of a cone (cf. (\ref{metric})), which in view of
(\ref{2.1}) gives
\begin{equation}
\omega^{12}_r=\omega^{21}_r= 0,\quad \omega^{12}_{\varphi}=
-\omega^{21}_{\varphi}=
1-\alpha.
\label{eq:3.2}\end{equation}
At $\nu=0$ it goes over to a flat one.
A topologically nontrivial gauge field (\ref{0.0})
results in a conical singularity of the spin connection.
The flux
$$\oint_{C} \omega^{12}_{\varphi}d\varphi=2\pi\nu\ne 0.$$
represents a "net" effect produced by a disclination on the moving electrons.
We thus show that the gauge-field approach within the linear approximation exactly coincides with that
provided by the "cut-and-glue" procedure.

\section{Flexible surface}

As a matter of fact, a cone with a point-like apex is mathematical abstraction since in a real situation the media has a
finite stiffness, which would inevitably result in a certain smearing of a conical singularity.
Therefore, a proper description of the disclination implies a smooth deformation of the metric
and at the same time  one has to preserve a conical behavior far away from the origin.
Although such a surface can effectively be approximated by a hyperboloid, we show
now that one cannot incorporate finite elasticity into the theory by simply replacing a cone by a smooth surface that
asymptotically approaches a cone far away from the origin. This would simply kill the defect.

To illustrate this, consider an upper half of a hyperboloid as an
embedding
$$(\chi,\varphi)\to (a\,{\sinh\chi}\cos\varphi,a\,{\sinh\chi}\sin\varphi,
c\,\cosh\chi), \quad 0\le\chi<\infty, 0\le\varphi<2\pi,$$
The components of the induced metric can be written as
\begin{equation}
g_{\chi\chi}=a^2\cosh^2\chi+c^2\sinh^2\chi,\quad g_{\varphi\varphi}=
a^2\sinh^2\chi,\quad g_{\varphi\chi}=g_{\chi\varphi}=0,
\label{4.1}\end{equation} which in view of (\ref{2.1}) gives for the
spin connection coefficients
\begin{equation}
\omega^{12}_{\chi}=\omega^{21}_{\chi}= 0,\quad \omega^{12}_{\varphi}=
-\omega^{21}_{\varphi}=
\left[1-\frac{a\,\cosh\chi}{\sqrt{g_{\chi\chi}}}\right]=:\omega(\chi).
\label{4.4}\end{equation}
The spin connection  in the spinor $SO(2)$ representation becomes
\begin{equation}
\Omega_{\varphi}=i\omega\sigma^3.
\label{4.5}\end{equation}
Since $\omega(\chi)$ goes to zero as $\chi\to 0$ a
circulation of that field over a loop encircling the origin gives a flux
which tends to zero as the counter shrinks to zero,
$$\lim_{\epsilon\to 0}\oint_{C_{\epsilon}}\omega^{12}=0,$$
where $C_{\epsilon}$ stands for a closed counter which encloses a small area $\sim\epsilon^2$ around the origin.
This equation implies that there is no a topologically nontrivial part in the flux.
It is therefore clear that one should work out some other way to explicitly accomodate elastic deformations in
the "cut-and-glue" procedure that would preserve a conical singularity at the origin.

We  show now that the gauge-theory appoach provides the necessary
framework. Within this theory the nontrivial flux is kept intact and
at the same time the elastic deformations are allowed for. Since a
plane can smoothly be deformed into a hyperboloid, a
hyperboloid-type smearing naturally emerges due to the elastic
deformations of the elastic plane. Actually, the proper metric could
be obtained from the exact solution of self-consistent equations in
the continuum theory of buckled membranes (see (4.10)
in~\cite{nelson} and (25) in~\cite{jpa99}). However, this very
complicated problem is still unsolved. Instead, we suggest to fulfil
these requirements by applying restrictions on the parameters of the
hyperboloid. Namely, the parameter $c$ must be proportional to
$\nu^{\eta}$ while the parameter $a$ should depend on both the
Young's modulus $K_0$ and the bending rigidity $\kappa$. In fact,
the first condition comes from a trial solution away from the
disclination core found in~\cite{nelson} which reads
$c\sim\nu^{1/2}$. The parameter $a(K_0,\kappa)$ must meet the
condition $a(K_0,\kappa)\rightarrow\infty$ at
$K_0\rightarrow\infty$. Indeed, the intrinsic curvature of the
hyperboloid reads
\begin{equation}
K = \frac{c^2}{(a^2\cosh^2\chi+c^2\sinh^2\chi)^2}
\label{curvature}
\end{equation}
which vanishes at $K_0\rightarrow\infty$ as it should be in the
inextensional limit. At the same time, a disclination defect sitting
at the origin is taken care of by the gauge field (\ref{0.0}). In
accordance with (\ref{2.0}) this field induces the following
explicit changes in the geometry of the hyperboloid:
\begin{equation}
g_{\varphi\varphi}= a^2\alpha^2\sinh^2\chi,
\label{4.6}
\end{equation}
and
\begin{equation}
\omega^{12}_{\varphi}=
-\omega^{21}_{\varphi}=
\left[1-\frac{a\alpha\,\cosh\chi}{\sqrt{g_{\chi\chi}}}\right]=:\omega_{\alpha}(\chi).
\label{4.9}\end{equation}
Here again $\alpha=1-\nu.$
Since $$\omega_{\alpha}(\chi)\to 1-\alpha, \quad\chi\to 0,$$
this spin connection term in contrast with (\ref{4.4}) contains
a topologically nontrivial part that gives rise to a fixed
flux,
$$\lim_{\epsilon\to 0}\oint_{C_{\epsilon}}\omega^{12}=2\pi\nu.$$
We thus finally get the smoothed apex, the cone-like asymptotic at
large distances and the unremovable conical singularity at the
disclination line. It is known that in case a spin connection
contains an $SO(2)$ piece with nontrivial flux, that field cannot be
eliminated under any smooth deformation of the underlying metric
(see, e.g., \cite{witten}). Within our approach this simply means
that a nontrivial contribution to the spin connection which comes
from the gauge field $W$ survives any smooth elastic deformations of
the media. Notice also that at large distances we expect only small
deviations from the cone resulting from the "cut and glue"
procedure, so that the physically reasonable restriction is
$\lambda=c/a\ll 1$. In other words, we restrict our consideration to
materials with high $K_0$.

In $2D$ the Dirac matrices can be chosen to be the Pauli matrices:
$\gamma^1=-\sigma^2, \gamma^2=\sigma^1$. The Dirac operator then
reads
\begin{eqnarray}
\hat{\cal D}=
\left[\matrix{0 & e^{-i\varphi}
\left(-\frac{\partial_{\chi}}
{\sqrt{g_{\chi\chi}}}+\frac{1}{a\alpha\sinh\chi}(i\partial_{\varphi}+\frac{1}{2}\omega_{\alpha}(\chi)
)\right)\cr
e^{i\varphi}
\left(\frac{\partial_{\chi}}
{\sqrt{g_{\chi\chi}}}+\frac{1}{a\alpha\sinh\chi}(i\partial_{\varphi}-\frac{1}{2}\omega_{\alpha}(\chi)
)\right)&0}
\right].
\label{4.10}\end{eqnarray}
At $\alpha=1\,(\nu=0)$ it reduces to a flat one.
For nonzero $\alpha$ and $K_0\rightarrow\infty$, the parameter $\lambda\rightarrow 0$ and
one obtains the Dirac operator on a cone.

The eigenfunctions to (\ref{4.10}) are classified with respect to the
eigenvalues of $J_z=j+1/2,\,\, j=0,\pm 1,\pm 2,...,$ and are to be taken in the
form
\begin{equation}
\psi =
\left(u(r)e^{i\varphi j}\atop{v(r)e^{i\varphi (j+1)}}\right).
\label{eq:1.5}\end{equation}
The substitution $$\tilde\psi={\psi}\sqrt{\sinh\chi}$$
reduces the eigenvalue problem (\ref{4.10}) to
\begin{eqnarray}
\partial_{\chi}\tilde u- \tilde{j}\sqrt{\coth^2\chi+\lambda^2}\,\tilde u=
\tilde{E}\tilde v, \nonumber \\
-\partial_{\chi}\tilde v
- \tilde{j}\sqrt{\coth^2\chi+\lambda^2}\,\tilde v=\tilde{E}\tilde u,
\label{eq:2.6}\end{eqnarray}
where $\tilde{E}=\sqrt{g_{\chi\chi}}\,E$  and $\tilde j=(j+1/2)/\alpha$.

One of the most interesting problems associated with Dirac fermions
in disclinated elastic media is the manifestation of the topological
effects. For various graphene surfaces these issues were discussed
in~\cite{jose92,jose93,lammert,jetplet00,jetplet01,pachos2,ando,jackiw,zaanen,sitenko}.
Let us study the influence of the smoothed cap on the emergence of
the zero-mode states in flexible disclinated materials. A general
solution to (\ref{eq:2.6}) at $E=0$ is found to be
\begin{eqnarray}
\tilde u_0(\chi)=A\left [\left (k\cosh\chi+\Delta\right )^{2k}
\frac{\Delta-\cosh\chi}{\Delta+\cosh\chi}\right]^{\frac{\tilde j}{2}}, \nonumber \\
\tilde v_0(\chi)=A\left [\left (k\cosh\chi+\Delta\right )^{2k}
\frac{\Delta-\cosh\chi}{\Delta+\cosh\chi}\right]^{-\frac{\tilde j}{2}}
\end{eqnarray}
where $k=\sqrt{1+\lambda^2}$, $\Delta=\sqrt{1+k^2\sinh^2\chi}$. The
normalization conditions read as follows: $ -1/2< \tilde j<-1/2k$
for $u_0(\chi)$ and $1/2k<\tilde j< 1/2$ for $v_0(\chi)$
(see~\cite{jetp}). As a result, at small $\lambda$, which is of
interest here, there are no normalized solutions. This means that in
stiff materials smoothing has no marked effect on the existence of
zero modes. The situation drastically changes in the presence of the
uniform magnetic field directed along the $z$-axis. In this case,
one of the modes (either $\tilde u(\chi)$ or $\tilde v(\chi)$)
becomes normalizable and there exists a true zero-energy state.
Therefore, one can expect "switching-like" effects governed by the
magnetic field. Studies of this problem are now in progress. Notice
that when $\lambda\rightarrow 0$ the wave functions $u_0(\chi)$ and
$v_0(\chi)$ vanish and we arrive at another class of solutions
typical for a true cone (see, e.g.,~\cite{lammert}).

\section{Conclusion}

In conclusion, we have presented a general approach based on the
gauge-theory and geometrical consideration which allows us to
describe Dirac fermions on flexible surfaces in the presence of a
disclination. Our model takes into account a 'seamless' smearing
of the conical singularity in the curvature thus avoiding the
evident problem at the core radius in (\ref{coremetric}). The
elasticity of the surface affects the embedding which is chosen to
be a plane in the inextentional limit and a hyperboloid-type
surface for a flexible material. Within our approach both the
elastic deformations and the topologically nontrivial gauge field
contribute to the induced metric, which in turn affects the spin
connection. It is in this way that Dirac fermions are affected by
the topological disclination defects. This approach provides a new
insight into disclination theory in a curved $2D$ background in
the presence of electrons and may therefore reveal some novel
physical phenomena. In order to apply our consideration to the
graphene-based curved materials, one has to take into account an
additional non-Abelian gauge field which is responsible for
coupling of two Fermi points. This study is now in progress and
the results will be published elsewhere.

This work has been supported by the Russian Foundation for Basic
Research under grant No. 08-02-01027.

\end{document}